\documentclass[conference,a4paper]{IEEEtran}
\usepackage{times}

\usepackage{graphicx,color}
\usepackage{cite}
\usepackage{amsmath}
\usepackage[breaklinks=true,colorlinks=true,backref,pagebackref]{hyperref}
\begin{document}

% Explicit linebreaks \\ can be added in the title. Do not put math or
% special symbols in the title.
\title{Generally Covariant Maxwell Theory for Media with a Local
  Response: Progress since 2000}

\author{\IEEEauthorblockN{%
    Friedrich W.\ Hehl\IEEEauthorrefmark{1}, Yakov
    Itin\IEEEauthorrefmark{2}, and Yuri N.\
    Obukhov\IEEEauthorrefmark{3}} \medskip
  \IEEEauthorblockA{\IEEEauthorrefmark{1} Univ. of Cologne, Germany,
    and Univ. of Missouri-Colombia, USA; e-mail:
    hehl@thp.uni-koeln.de}
  \IEEEauthorblockA{\IEEEauthorrefmark{2}Jerusalem College of
    Technology and Hebrew Univ.\ Jerusalem,
    Israel; email: itin@math.huji.ac.il}
  \IEEEauthorblockA{\IEEEauthorrefmark{3}%
    Nuclear Safety Institute, Russian Academy of Sciences, Moscow,
    Russia; email: obukhov@ibrae.ac.ru}}

\maketitle

% Do not use any symbols, special characters or math in the abstract.
\begin{abstract}
  In the recent decades, it became more and more popular for
  engineers, physicists, and mathematicians alike to put the Maxwell
  equations into a generally covariant form. This is particularly
  useful for understanding the fundamental structure of
  electrodynamics (conservation of electric charge and magnetic
  flux). Moreover, it is ideally suited for applying it to media with
  local (and mainly linear) response behavior. We try to collect the
  new knowledge that grew out of this development. We would like to
  ask the participants of EMTS 2016 to inform us of work that we may
  have overlooked in our review.

  28 March 2016, {\it file
    EspooHehlLocLinMax$\underline{\hspace{5pt}}${10}.tex}
\end{abstract}
\medskip
%\section{Introduction}

\begin{figure}[t]
    \centering
    \includegraphics[height=64.54mm]{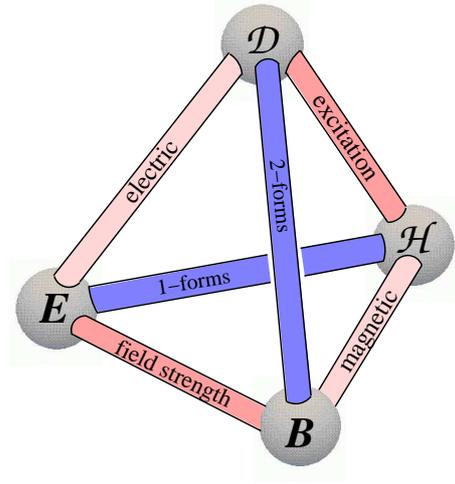}
    \caption{The electromagnetic field: The interrelation between the
      excitation $H=({\cal D}, \cal{H})$ and the field strength
      $F=(E,B)$.}
    \label{fig:tetrahedron}
\end{figure}

\begin{figure}[t]
    \centering
    \includegraphics[width=10cm%68.7mm
]{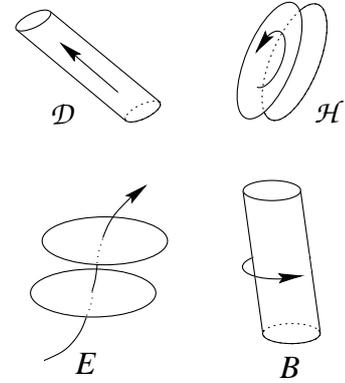}
    \caption{The Faraday-Schouten pictograms of the electromagnetic
      field in 3-dimensional space. The 1-forms are depicted by 2
      parallel planes, the 2-forms by flux tubes.}
    \label{fig:picto}
\end{figure}

Premetric classical electrodynamics has been consistently formulated
in 2003 by Hehl and Obukhov \cite{Birkbook}; earlier references can be
found in that book. We may add O'Dell \cite{O'Dell} and Serdyukov et
al.\ \cite{Serdyukov:2001}, which we had overlooked at the time. The
premetric framework consists of the generally covariant Maxwell
equations in the 4-dimensional or the (1+3)-decomposed version, see
Figs.\ \ref{fig:tetrahedron} and \ref{fig:picto}:
\begin{align}
dH &= J\>\,\begin{cases}\>\, \underline{d}\,{\cal
      D}\>=\rho\,,\\ \quad\,
    \dot{{\cal D}}\,=\underline{d}\,{\cal H}-j\,; %\hspace{-18pt}
\end{cases} \label{evol1}% \\
\end{align}\vspace{-9pt}\begin{align}
\hspace{-7pt}dF &= 0\>\,\,\begin{cases} \>\,\underline{d}\,B\>=0\,
   ,\\ 
    \quad\,\dot{B}\,=-\underline{d}\,E\,.%\hspace{-18pt}
\end{cases} \label{evol2}
\end{align}
If the medium has a local and linear response behavior, the excitation
$H=({\cal H},{\cal D})$ is local and linear in the field strength
$F=(E,B)$. In components we have
\begin{equation}
\check{H}^{ij}=\frac 12\chi^{ijkl}\,F_{kl}\quad\text{with}\quad
\chi^{ijkl}=-\chi^{jikl}=-\chi^{ijlk}\,.
\end{equation}
Here $i,j,...=0,1,2,3$ and
$\check{H}^{ij}:=(1/2)\epsilon^{ijkl}H_{kl}$. The response tensor
density $\chi$ has 36 independent components. It can be decomposed into
three irreducible pieces with $36=20+15+1$ components,
respectively. Split in (1+3) dimensions, with $a,b,..=1,2,3$, we have
(for details, see \cite{Birkbook,Hehl:2007ut})
\begin{equation*}
  {D}^a=\left( \varepsilon^{ab}- \,\epsilon^{abc}\,n_c \right)E_b\,
  +\left(\gamma^a{}_b + s_b{}^a - \delta_b^a s_c{}^c\right) {B}^b +
  {\alpha}\,B^a \,,
\end{equation*}
\begin{equation*}
  {H}_a=\left( \mu_{ab}^{-1}- {\epsilon}_{abc}m^c \right) {B}^b 
  +\left(- \gamma^b{}_a +s_a{}^b - \delta_a^b s_c{}^c\right)E_b 
  -{\alpha}\,E_a\,.\label{explicit''}
\end{equation*}

Up to here, the metric of spacetime, that is, the gravitational
potential, did not enter anywhere. We have a premetric framework.  The
different irreducible parts split as follows (see
\cite{Lange:2015zja}):
\begin{align*}
20&\:\text{components of principal part:}\;6\;\varepsilon^{ab}\,,\;
6\;\mu^{-1}_{ab}\,,\;8\;\gamma^a{}_b\,;\\
15&\;\text{components of skewon part:}\;9\;s_a{}^b\,,\;3\;n_c\,,\;3\;m^c\,;\\
1&\;\text{component of axion part:  }\alpha\,.
\end{align*}\vspace{-5pt}


Progress has been achieved since then by different groups, see the
more recent books of Delphenich \cite{Delphenich:2009}, Favaro
\cite{Favaro} (Ph.D.\ thesis), Lindell \cite{Ismobook,Ismo}, Raab \&
de Lange \cite{Raab:2005}, Russer \cite{Russer}, and Scheck
\cite{Scheck}, see also Cabral \& Lobo \cite{Cabral:2016yxh}. In
particular Scheck underlines that polarizability and magnetization
(and thus $\cal D$ and $\cal H$) are microscopically defined, ``i.e.\
for a single atom or even an elementary particle.''  \vspace{1pt}

Premetric electrodynamics turned out to be a lively and active
subject. Mainly the following aspects were developed (we order them
roughly chronologically):
\begin{enumerate}

\item\label{magneticI} {\sl Magnetic charge is alien to the Maxwellian
    structure:} A {magnetic charge\/} can be mathematically included
  in premetric electrodynamics in a consistent way, see Edelen
  \cite{Edelen}, Kaiser \cite{Kaiser}, and Hehl \& Obukhov
  \cite{magnetic}.\vspace{2pt}

  As shown in \cite{Birkbook}, the structure of the Maxwell equations
  can be understood as arising from (i) electric charge conservation
  $dJ=0$ and (ii) magnetic flux conservation $dF=0$.  This distinction
  between the electric and magnetic counterparts of the theory can be
  viewed as a reflection of the topology properties of the underlying
  manifold: The electric current $J$ and the corresponding excitation
  $H$ are twisted forms, whereas the field strength $F$ is
  untwisted. If we modified the electromagnetic structure, following
  Dirac, by introducing a magnetic current $K$, with $dF=K$, the
  Maxwell equations would be ``symmetrized,'' but the distinction
  between the two basic fields $H$ and $F$ would be destroyed.

\item\label{skewon} {\sl Skewon part of the electromagnetic response
    tensor:} The general concept of the skewon part of the response
  tensor was introduced in \cite{skewonMex}, for its relation to
  dissipation, see \cite{Birkbook}. The effects of the skewon on light
  propagation were studied by Itin, Ni, Obukhov, et al. \cite{skewon},
  \cite{Itin:2013ica}, \cite{Itin:2015tdi}, \cite{Mei:2014iaa},
  \cite{Ni:2014qfa}, \cite{Itin:2014gba}, \cite{Itin:2014rwa},
  \cite{Ni:2013uwa}. A theory of the skewon in interaction with the
  gravitational field was provided in \cite{Hehl:2005xu}.

\item\label{QHEI} {\sl Quantum Hall effect is independent of the
    external gravitational field:} The {QHE} can be phenomenologically
  described in a $(1+2)$-dimensional premetric electrodynamic
  framework; also the local, linear response tensor turns out to be
  premetric. Therefore, the QHE cannot depend on an external
  gravitational field, see %Hehl, Obukhov, and Rosenow
  \cite{Rosenow}.

\item\label{nonminimalI} {\sl Electromagnetism can couple
    to a possible Cartan torsion of spacetime only nonminimally:}
  Solanki, Preuss, et al.\ \cite{Solanki,Preuss} pointed out that a
  non-minimal coupling of gravity to electromagnetism, in particular
  to the torsion field, is possible, see also Hehl and Obukhov
  \cite{gyros}, Rubilar et al.\ \cite{torsion1}, and Itin et al.\
  \cite{Itin:2003jp,torsion2}.

\begin{figure}\label{fig9}
\includegraphics[width=8cm,height=7.5cm]{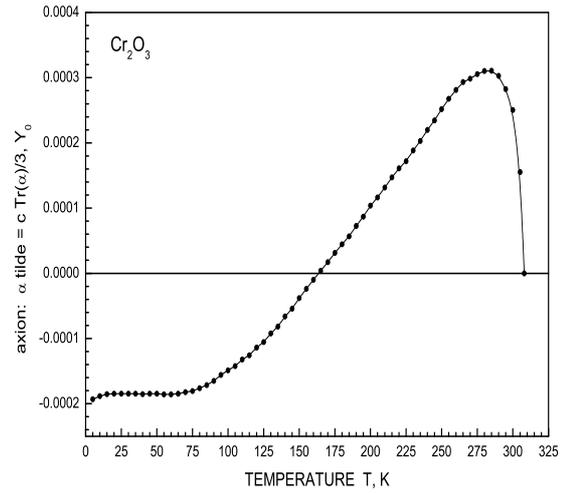}
\caption{The pseudoscalar or axion piece ${\alpha}$ of the response
  tensor $\chi^{ijkl}$ of Chromium Sesquioxide Cr$_2$O$_3$; in units
  of $Y_0=1/Z_0$ as a function of the temperature $T$ in {\it kelvin}.
  Here $Z_0$ is the vacuum impedance which, in SI, is $\approx
  377\hspace{-2pt}$ {\it ohm}.} \label{fig:chromium}
\end{figure}

\item\label{signatureI} {\sl Signature of the metric can
    be derived from electrodynamics:} In \cite{Birkbook} it has been
  pointed out that the signature of the metric can be derived from the
  Lenz rule and the positivity of the energy of the electromagnetic
  field. The formalism has been straightened out, improved, and
  enriched by Itin et al.\ \cite{Annals,ItinFriedman}.

\item\label{Tellegen} {\sl Four incarnations of the Tellegen gyrator:}
  The Tellegen gyrator \cite{Tellegen1948,Tellegen1956/7,Kupf,Russer}
  is an element in linear and passive network theory, which
  ``rotates'' currents $I$ into voltages $V$. Similarly, (i) the axion
  part $\alpha$ of the electromagnetic response tensor rotates
  excitations $({\cal H},{\cal D})$ into field strengths $(B,E)$
  \cite[Eqs.(D.1.112/3)]{Birkbook}, and so does (ii) the PEMC (perfect
  electromagnetic conductor) of Lindell \& Sihvola
  \cite{LindSihv2004a,LindSihv2004b}, see \cite[Eq.(7.72)]{Ismo}.
  (iii), in the QHE, the current/charge $(j,\rho)$ is rotated into the
  field strength $(E,B)$, see
  \cite[Eq.(B.4.64)]{Birkbook}.\footnote{In spite of the last
    mentioned fact, one cannot build a gyrator out of the QHE, as
    pointed out to us by von Klitzing \cite{vonKlitzing}.}  (iv) The
  hypothetical axion particle of elementary particle physics, see
  Wilczek \cite{Wilczek:1987mv} and Weinberg \cite[pp.\
  458--461]{Weinberg} has the same behavior as (i) and (ii). In the
  gyrator and additionally, in its four incarnations, the P-symmetry
  (parity) and the T-symmetry (time) are always violated. It has been
  experimentally confirmed \cite{Hehl:2007ut}, see
  Fig.\ref{fig:chromium}, that in single crystals of Chromium
  Sesquioxide Cr$_2$O$_3$ at low temperatures the axion part is
  non-vanishing and maximally of the order of 10$^{-3}$ $\times$
  vacuum admittance. This falsifies the so-called Post constraint.

\item\label{CFJI} {\sl Axion and light propagation:} Ni \cite{Ni73},
  \cite{Ni77} proposed, as a device for violation the equivalence
  principle, to modify classical electrodynamics by introducing an
  axion field. Carroll et al.\ \cite{CFJ} used an axion field with a
  time-like gradient for violating Lorentz and parity invariance.
  This model was extended and reformulated in the premetric formalism
  by Itin \cite{Yakov1}, \cite{Itin:2007cv},
  \cite{Itin:2007wz}. Various theoretical issues of this model were
  studied by Balakin \cite{Balakin:2012up}, and
  Noble\cite{Noble:2008zza}. Recently Ni \cite{Ni:2015poa},
  \cite{Ni:2014cca} provided constraints on axion and dilaton from
  polarized/unpolarized laboratory/astrophysical/cosmic
  experiments/observations.

\item\label{birefringenceI} {\sl Forbid birefringence in vacuo and
    find the light cone:} Wave propagation in electrodynamics with a
  local and linear response law exhibits the phenomenon of
  birefringence. In the geometric optics approximation, this is
  manifest in the fourth order Fresnel equation for the wave covector.
  Such a picture is typical for complex (anisotropic, magnetoelectric,
  moving) media. However, it is natural to assume that the physical
  vacuum is a non-birefringent continuum. This no-birefringence
  condition reduces the general Fresnel equation to the light cone
  \cite{lightcone,forerunner}; thus, the spacetime metric is recovered
  from the electromagnetic response tensor. More recent developments
  can be found in \cite{G1,G2,G3}.

\item\label{dimensionI} {\sl Dimensional analysis of physical
    quantities:} In \cite{Birkbook} we analyzed electrodynamics in
  terms of a consistent theory of physical dimension. In particular
  the operational verification of the physical quantities were at
  center stage. In \cite{Okun}, we extended these considerations in
  the context of a discussion with Okun. The vacuum admittance
  $\sqrt{\varepsilon_0 / \mu_0}=e^2/(2h\alpha)$ can be expressed in
  terms of the fine structure constant $\alpha$, the elementary charge
  $e$ and Planck's constant $h$. The possible time variability of the
  vacuum impedance and the fine structure constant were discussed by
  Tobar \cite{Tobar} and us.

\item\label{BeigI} {\sl New initial value formulation of Maxwell's
    theory by Perlick:} It was studied by Perlick
  \cite{Perlick:2010ya} for the premetric version.  He derived several
  conditions for the evolution equations to be hyperbolic, strongly
  hyperbolic, or symmetric hyperbolic. In particular, he
  characterized all response laws for which the evolution equations
  are symmetric hyperbolic. The latter property is sufficient, but not
  necessary, for well-posedness of the initial-value problem. Symbol
  and {\it hyperbolic polynomials\/} of the wave equation and their
  relation to premetric electrodynamics were discussed by Beig
  \cite{Beig}.  Itin \cite{Itin:2014wea} derived premetric covariant
  jump conditions for the initial value hypersurface, the light cone,
  and a boundary between two media.

\begin{center}
\begin{figure}\label{fig9}
\includegraphics[width=11cm,height=11cm]{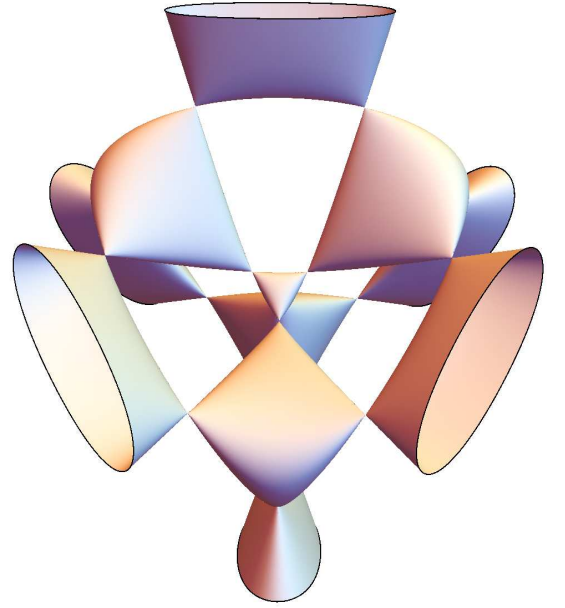}
\caption{The Fresnel surface for a metamaterial with the moduli as
  displayed in equation (5). This surface turns out to be a Kummer
  surface of algebraic geometry.} \label{fig:kummer}
\end{figure}
\end{center}

\item\label{Kummer} {\sl The Kummer tensor of electrodynamics, Fresnel
    wave surfaces unmasked as Kummer surfaces:} We found a fourth rank
  Kummer tensor density \cite{Baekler:2014kha} that is cubic in the
  response tensor $\chi$,
    \begin{equation} {\cal
        K}^{ijkl}:=\chi^{aibj}\,^\diamond\chi^\diamond_{acbd}\,\chi^{ckdl}\,,
    \end{equation} 
    where $^\diamond$ denotes a premetric duality operator. The
    Fresnel surfaces are defined by ${\cal K}^{ijkl}\,q_iq_jq_kq_l=0$;
    here $q_i$ is the wave covector of the propagating
    light. Following earlier ideas of Bateman, Favaro et al.\
    \cite{Favaro:2015jxa} were able to show that, for vanishing skewon
    part, the Fresnel surfaces turn out to be Kummer surfaces with up
    to 16 singular points. By a suitable choice of the electromagnetic
    moduli of a hypothetical metamaterial according to
    \cite{Favaro:2015jxa},
    \begin{align}
      \varepsilon^{ab}&=\textstyle\frac{1}{4}\varepsilon_0\,\mbox{diag}
      \bigl(-1-\sqrt{3},-1-\sqrt{3},-4+2\sqrt{3}\bigr),\nonumber\\
      \mu^{-1}_{ab}&=\textstyle\frac{1}{4}\mu_0^{-1}\,\mbox{diag}
      \bigl(1+\sqrt{3},1+\sqrt{3},4-2\sqrt{3}\bigr),%\nonumber
\\
      \gamma^{a}{}_{b}&=\textstyle\frac{1}{4}\sqrt{{\varepsilon_0}{\mu_0^{-1}}}\;
      \mbox{diag} \bigl(3+\sqrt{3},-3-\sqrt{3},0\bigr),\nonumber
\end{align}
it was possible to create such a surface, see Fig.\ \ref{fig:kummer}.

Russer pointed out to us \cite{RusserEmail} that the 16-fold symmetry
of the Fresnel-Kummer surface reminds him of a 16-fold solution
manifold in the transmission-line matrix (TLM) method, see
\cite[Sec.14.3]{Russer} and \cite{Russer+Russer}. Since both phenomena
refer to the propagation of electromagnetic signals, there may very
well exist an interrelationship.

\item\label{factor} {\sl Possible factorizations of the Fresnel
    equation (FE):} Which media give rise to such simplifications?
  Four cases were solved: (i) The FE cannot be factorized in
  polynomials of lower degree. Materials with anisotropic permittivity
  and permeability, e.g., display this behavior \cite{H1}. (ii) The FE
  factorizes in irreducible quadratic polynomials. For research on the
  situation where the two factors are distinct and have Lorentzian
  signature, see \cite{H2,H3,H4}. Light then propagates along two
  distinct light cones. When the two quadratic factors are the same,
  up to a constant, a single light cone is obtained, so that there is
  no birefringence, see item 8). (iii) The FE factorizes in four
  linear polynomials. Extreme Magneto-Electric (EME) materials
  \cite{H5} are known to have this property. (iv) The FE vanishes
  trivially, i.e., it is satisfied by any wave covector
  $q_{i}=(\omega,-k_a)$. Media corresponding to this scenario are
  described in \cite{H6}.

\item\label{PhotonInSkew} {\sl Photon propagator in premetric
    electrodynamics:} The Green tensor is the photon propagator in
  momentum space. It is important in quantum field theoretical
  extension of electrodynamics. It was derived and applied to axion
  and skewon modified electrodynamics by Itin \cite{Itin:2007av},
  \cite{Itin:2007cv}, \cite{Itin:2015tdi}. In a similar way, Andrianov
  et al.\ \cite{Andrianov:2011wj} studied the propagation of photons
  and massive vector mesons in the presence of a medium, which
  violates Lorentz and a CPT invariance. Pfeifer and Siemssen
  \cite{Pfeifer:2016har} applied the premetric photon propagator for
  studying the causal structure and the quantization problems in
  electrodynamic models with linear response.

\item\label{acceleration} {\sl Accelerated reference frames in
    coordinate free form:} Auchmann and Kurz \cite{Auchmann:2014cqa},
  by using a fiber bundle formalism, developed a general theory for
  observers by splitting 4-dimensional electrodynamics in a covariant
  way into time and space. This is particularly useful for
  interpreting experiments in accelerated motion.

\item\label{dissipation} {\sl Energy-momentum tensor of the
    electromagnetic field \cite{YuriEMTS2016,Ramos}:} The form of the
  energy-momentum tensor of the electromagnetic field in a medium with
  nontrivial electric and magnetic properties is a problem that dates
  back to the century-old work of Minkowski and Abraham.  The
  Minkowski vs.\ Abraham controversy can be resolved by the crucial
  observation that the electromagnetic field in a medium is an open
  physical system. The closed system of interacting electromagnetism
  plus matter is characterized by the total energy-momentum tensor
  that is defined unambiguously within the Lagrange-Noether
  framework. Both Minkowski and Abraham tensors may arise under
  appropriate conditions for special splits of the total
  energy-momentum tensor.

\item\label{nonlinearI} {\sl Nonlinear electrodynamics:} So far, we
  only treated linear response laws. A nonlinear law naturally arises
  when the quantized electromagnetic field excites radiative vacuum
  corrections. Alternatively, one can consider a fundamentally
  nonlinear electrodynamics with a Born-Infeld type Lagrangian, see
  \cite{Birkbook}. Wave propagation in nonlinear models manifest
  itself typically in the birefringence property. An analysis of
  geometric optics in nonlinear electrodynamics reveals the product
  structure of the Fresnel wave surface, which gives rise to two
  optical metrics, see %Obukhov \& Rubilar 
\cite{Obukhov:2002xa}. It is
  also possible to formulate a Finsler electrodynamics in a premetric
  way, see %Delphenich 
\cite{DelphenichNonlinear} and %Itin et al.\
  \cite{Itin:2014uia}.
  
 \end{enumerate}

\bibliographystyle{IEEEtran}
\bibliography{IEEEabrv,IEEEexample}

\section*{Acknowledgments}

We are very grateful to Alberto Favaro (London) for many useful
discussions and detailed remarks. We'd also like to thank Gerry Kaiser
(Portland, OR) for a discussion on the possible existence of magnetic
charge.

\end{document}